\documentclass[graybox]{svmult}

\usepackage{mathptmx}
\usepackage{helvet}
\usepackage{courier}
\usepackage{makeidx}
\usepackage{type1cm}        

\usepackage{graphicx} 
\usepackage{multicol}
\usepackage[bottom]{footmisc}

\makeindex

\usepackage {amsmath}
\usepackage {txfonts}
\usepackage {hyperref}
\usepackage {url}
\usepackage {acronym}

\newcommand{\tn}{\textnormal}

\begin{document}

\title*{Coincidence searches of gravitational waves and short gamma-ray bursts}

\author{Andrea Maselli and Valeria Ferrari}

\institute{Andrea Maselli \at Dipartimento di Fisica, Universit\`a di Roma ``La Sapienza'' \& Sezione, 
INFN Roma1, P.A. Moro 5, 00185, Roma, Italy, \email{andrea.maselli@roma1.infn.it}
\and Valeria Ferrari \at Dipartimento di Fisica, Universit\`a di Roma ``La Sapienza'' \& Sezione, 
INFN Roma1, P.A. Moro 5, 00185, Roma, Italy, \email{valeria.ferrari@roma1.infn.it}}

\maketitle

\abstract{
Black-hole neutron-star coalescing binaries have been invoked as one of the 
most suitable scenario to explain the emission of short gamma-ray bursts. 
Indeed, if the black-hole which forms after the merger, is surrounded by a massive 
disk, neutrino annihilation processes may produce high-energy and collimated 
electromagnetic radiation. In this paper, we devise a new procedure, to  be used 
in the search for gravitational waves from black-hole-neutron-star binaries, to 
assign a probability that a detected gravitational signal is associated to the 
formation of an accreting disk, massive enough to power gamma-ray bursts. 
This method is based on two recently proposed semi-analytic fits, one reproducing 
the mass of the remnant disk surrounding the black hole as a function of some binary 
parameters, the second relating the neutron star compactness, with its tidal deformability. 
Our approach can be used in low-latency data analysis to restrict the parameter space 
searching for gravitational signals associated with short gamma-ray bursts, and to 
gain information on the dynamics of the coalescing system and on the neutron 
star equation of state.}
 

\section{Introduction}

Coalescing binary systems formed by neutron stars (NSs) and/or black 
holes (BHs) represent one of the most promising sources of gravitational 
waves (GWs) to be detected by interferometric detectors of second 
(AdvLIGO/Virgo) and third (the Einstein Telescope, ET) generation 
\cite{LIGOVirgo,ET1}. Moreover, these events have recently been proposed 
as a candidate for the central engine of short gamma-ray bursts (SGRB), 
provided the stellar-mass BH which forms after merging is surrounded by 
a hot and sufficiently massive accreting disk (see for instance \cite{LR07} 
and references therein). Since the electromagnetic emission is produced 
at large distance from the central engine, it does not give strong information 
on the source. In addition, the emission is beamed, and consequently these 
events may not be detected if one is looking in the wrong direction. 
Conversely, the gravitational wave (GW) emission is not beamed, and 
exhibits a characteristic waveform (the chirp), which should allow a 
non-ambiguous identification of the source. 
GRBs are characterized by a prompt emission, which lasts a few seconds, 
and an afterglow, whose duration ranges from hours to days. 
Thus, gravitational wave detection may be used to trigger the afterglow 
search of  GRBs that have not been detected by the on-axis prompt 
observation and to validate  the ``jet model'' of SGRB.
Or, alternatively, the observation of a SGRB may be used as a trigger
to search for a coincident GW signal. Indeed, this kind of search has
already been done in the data of LIGO and Virgo \cite{Abadieb,Abadiec}.

However, not all coalescences of compact bodies produce a black hole 
with an accreting disk sufficiently massive to power a SGRB: it is therefore 
crucial to devise a strategy to extract those having the largest probability to 
produce a SGRB. This is one of the purposes of this work.
The LIGO-Virgo Collaboration has recently developed a plausible observing 
schedule, according to which within this decade the advanced 
detectors, operating under appropriate conditions, will be able to determine the 
sky location of a source within $5$ and $20$ deg$^2$ \cite{LIGOVirgoplan}. 
Given the cost of spanning this quite large region of sky  to search for a 
coincident SGRB with electromagnetic detectors, indications on whether a 
detected signal is likely to be associated with a SGRB are valuable information. 

The procedure we propose has several applications. It can be used in the data 
analysis of future detectors  (i) to gain information on the range of parameters 
that are more useful to span in the low-latency search for GWs emitted by BH-NS 
sources \cite{Abadiea},  (ii) for an externally triggered search for GW coalescence 
signals following GRB observations \cite{Abadieb,Abadiec}, and (iii) when the 
binary parameters are measured with sufficient accuracy and in a sufficiently 
short time to allow for an electromagnetic follow-up to search for off-axis GRB 
afterglows. 


\section{Selecting candidates for gamma-ray bursts emission}

In the last years a large number of numerical studies of BH-NS coalescence, 
have allowed to derive two interesting fits. The first \cite{F12} gives the mass of 
the accretion disk, $M_{\tn{rem}}$, as a function of the the NS compactness
${\cal C}=M_{\tn{NS}}/R_\tn{NS}$, where $M_{\tn{NS}}$ and $R_\tn{NS}$
are the NS gravitational mass and its radius,
the dimensionless BH spin, $\chi_{\tn{BH}}\in[-1,1]$, and the mass ratio
$q=M_{\tn{BH}}/M_{\tn{NS}}$:
\begin{equation}\label{torusFit}
\frac{M_{\tn{rem}}}{M_{\tn{NS}}^{b}}=K_1 (3q)^{1/3}(1-2{\cal C})
-K_2 q\ {\cal C}\ R_{\tn{ISCO}}\ .
\end{equation}
Here $M^\tn{b}_\tn{NS}$  is the NS baryonic mass which,
following \cite{GPRTL13}, we assume to be  10\% larger than the 
NS gravitational mass;
$R_{\tn{ISCO}}$ is the radius of the innermost, stable circular orbit 
for a Kerr black hole \cite{BPT72}.
The two coefficients $K_1=0.288\pm0.011$ and $K_2=0.1248\pm 0.007$ have
been derived \cite{F12} through a least-square fit of the results of fully
relativistic numerical simulations \cite{KOST11,ELSB09,FDKSST12,FDKT11}.

$M_{\tn {rem}}$ is a key parameter in our study. Indeed, neutrino-antineutrino 
annihilation processes extract energy from the disk \cite{Piran04}, and several 
studies have shown that this process could supply the energy required to ignite 
a short gamma-ray burst, if $M_{\tn {rem}}\in(0.01\div0.05)M_{\tn{NS}}$ \cite{SLB13}. 
In the following we shall assume as a threshold for SGRB formation 
$M_{\tn{rem}}=0.01~M_{\tn{NS}}$. 
 
The second fit \cite{MCFGP13} is a universal relation between the NS 
compactness ${\cal C}$ and the tidal deformability $ \lambda_{2}=-Q_{ij}/C_{ij}$, 
where $Q_{ij}$ is the NS star traceless quadrupole tensor, and $C_{ij}$ is the 
tidal tensor,
\begin{equation}\label{Clambda} 
{\cal C}=0.371-3.9\times 10^{-2}\ln \bar{\lambda}
+1.056\times 10^{-3}(\ln\bar{\lambda})^2\quad\ , \quad \bar{\lambda}=\lambda_{2}/M_{\tn{NS}}^5\ .
\end{equation}
Hereafter, we shall denote by ${\cal C}_\lambda$ the NS compactness 
obtained from this fit.

\begin{figure}[ht]
\begin{center}
\includegraphics[width=5.5cm]{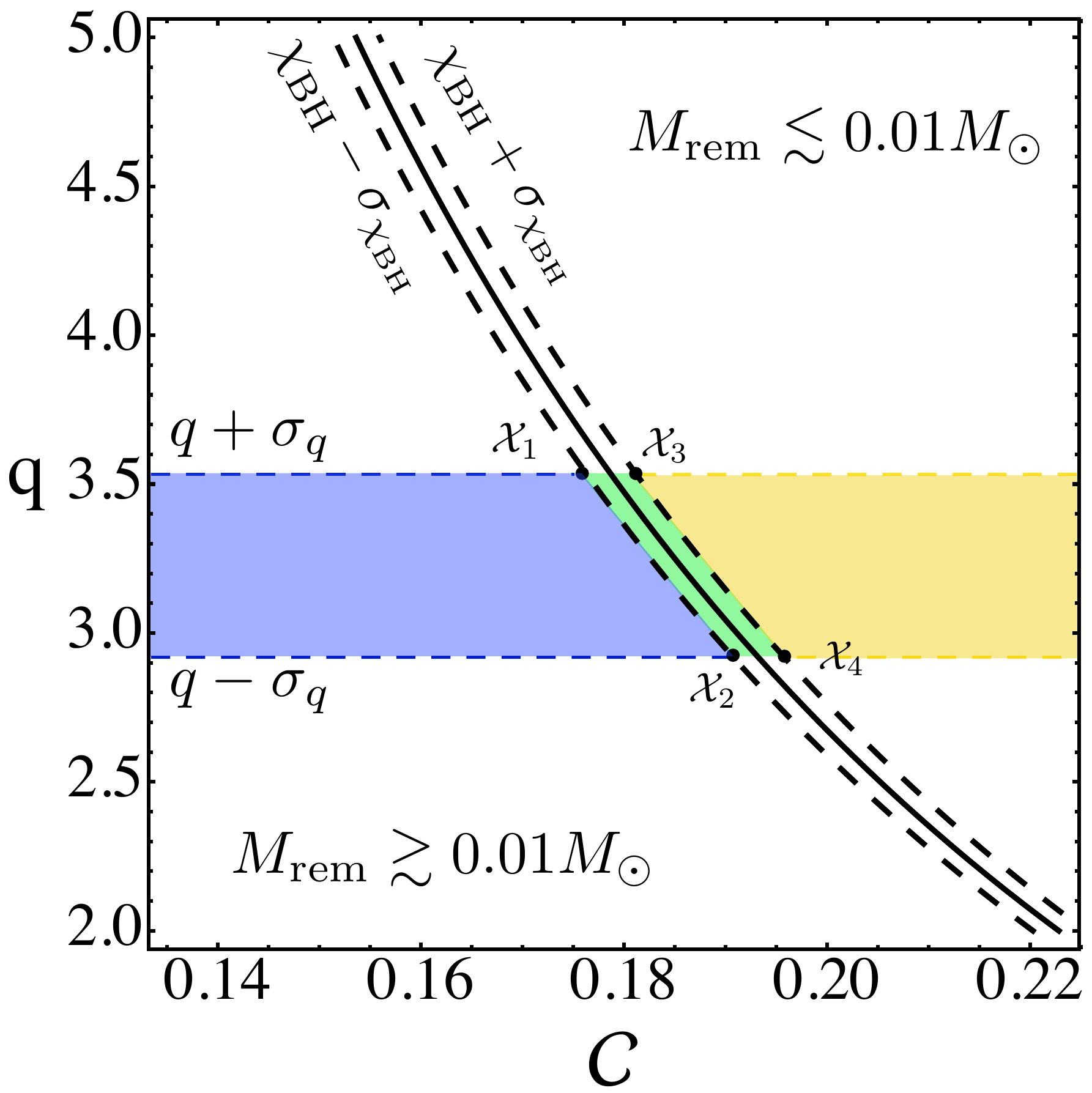}
\caption{Contour plot of the fit (\ref{torusFit}) in the $q$-${\cal C}$
plane, for $M_{\tn{NS}}=1.2M_{\odot}$, $\chi_{\tn{BH}}=0.5$ and
$M_{\tn{rem}} =0.01 M_{\odot}$. The fit  separates the region allowed
for SGRB ignition (below the fit curve) from the forbidden region (above
the fit). Given the measured values of $q\pm \sigma_q$ and
$\chi_{\tn{BH}}\pm\sigma_{\chi_{\tn{BH}}}$,
a detected signal can correspond to a NS with compactness 
${\cal C}$ which falls in one of the regions bounded by the dashed curves. 
Since ${\cal C}$ also comes with an error $\sigma_{\cal C}$,
in order to infer if  it can be associated with a SGRB, we need to
evaluate the probability $\tn{P}({\cal C} \leq {\cal
C}_{4})$ and $\tn{P}({\cal C} \leq {\cal C}_{1})$ (see text).
}
\label{FIG1}
\end{center}
\end{figure}

Let us now assume that the gravitational wave signal emitted in a BH-NS
coalescence is detected; a suitable  data analysis  allows us to find the
values of the mass-ratio $q=M_{\tn{BH}}/M_{\tn{NS}}$, of the chirp 
mass ${\cal M} = (M_{\tn{NS}}M_{\tn{BH}})^{3/5}/(M_{\tn{NS}}+M_{\tn{BH}})^{1/5}$, 
and of the black hole spin $\chi_{\tn{BH}}$, with the corresponding errors.
Knowing $q\pm \sigma_{q}$ and
$\chi_{\tn{BH}}\pm\sigma_{\chi_{\tn{BH}}}$, using the fit
(\ref{torusFit}) we can trace the plot of Fig.~\ref{FIG1} in the
$q-{\cal C}$ plane, for an assigned disk mass threshold, say
$M_{\tn{rem}} =0.01 M_{\tn{NS}}$.  This plot allows us to identify the
parameter region where a SGRB may occur, i.e., the region
$M_{\tn{rem}}\gtrsim 0.01 M_{\tn{NS}}$ (below the fit curve in the figure), 
and the forbidden region above the fit ($M_{\tn{rem}}\lesssim 0.01 M_{\tn{NS}}$).  
In addition, we
identify four points ${\cal X}_{1},\ldots{\cal X}_{4}$, which are the
intersection between the contour lines for
$\chi_{\tn{BH}}\pm\sigma_{\chi_{\tn{BH}}}$ and the horizontal lines $q\pm \sigma_{q}$.
Let us indicate as ${\cal C}_{1}, \ldots,{\cal C}_{4}$ the corresponding
values of the neutron star compactness. 
Since  the  fit (\ref{torusFit}) is monotonically decreasing, ${\cal
C}_{1}<{\cal C}_{2}<{\cal C}_{3}<{\cal C}_{4}$.  
At this stage we still cannot say whether the detected binary falls in the 
region allowed for the formation of a SGRB or not.
In order to get this information, we need to evaluate 
${\cal C}$. As discussed in \cite{DABV13,DNV12,Retal13,PROR11,MGF13,RMSUCF09}, 
Advanced LIGO/Virgo are expected to measure the gravitational wave phase with 
an accuracy sufficient to estimate the NS tidal deformability $\lambda_2$. Thus, using 
the fit (\ref{Clambda}), the neutron star compactness ${\cal C}_{\lambda}$ 
and the corresponding uncertainty $\sigma_{{\cal C}_{\lambda}}$ can be derived
(see \cite{MCFGP13} for details on how to compute the compactness error).

Knowing the parameters and their uncertainties, the probability that a SGRB is 
associated to the detected coalescence can now be evaluated. We assume that 
$(q,\cal{C}_{\lambda},\chi_\tn{{BH}})$ are described by a multivariate Gaussian 
distribution,
\begin{equation}\label{probab}
{\cal P}(q,{\cal C}_{\lambda},\chi_{\tn{BH}})=\frac{1}{(2\pi)^{3/2}
\vert\Sigma\vert^{1/2}}
\tn{exp}\left[-\frac{1}{2}\Delta^{\tn{T}}\Sigma^{-1}\Delta\right]\ ,
\end{equation}
where $\Delta=(\vec{x}-\vec{\mu})$, ~ $\vec{\mu}=(q,{\cal
C}_{\lambda},\chi_{\tn{BH}})$, and $\Sigma$ is the covariance matrix. Then, we 
define the maximum and minimum probability that the binary coalescence 
produces an accretion disk with mass over the threshold,  $\bar{M}_{\tn{rem}}$, 
as
\begin{eqnarray}\label{Pminmax}
\tn{P}_{\tn{MAX}}(M_{\tn{rem}}\gtrsim \bar{M}_{\tn{rem}} )
\equiv\tn{P}({\cal C}_{\lambda} \leq {\cal C}_{4})\ \ ,\quad
\tn{P}_{\tn{MIN}}(M_{\tn{rem}}\gtrsim \bar{M}_{\tn{rem}} )
\equiv\tn{P}({\cal C}_{\lambda} \leq {\cal C}_{1})\ ,
\end{eqnarray}
where $\tn{P}({\cal C}_{\lambda} \leq {\cal C}_{i})$
is the cumulative distribution of Eq. (\ref{probab}), which gives  
the probability that the measured compactness ${\cal C}_{\lambda}$, 
estimated through the fit (\ref{Clambda}), is smaller than an 
assigned value ${\cal C}_{i}$.

As an illustrative example,
we now evaluate the probability that a given BH-NS coalescing binary
produces a SGRB, assuming a set of equations of state for the NS matter
and evaluating the uncertainties on the relevant parameters using a
Fisher matrix approach.

\section{The uncertainties on the binary parameters}

The accuracy with which future interferometers will measure a set of binary
parameters $\boldsymbol\theta$ is estimated by comparing the
gravity-wave data stream with a set of theoretical templates.
For strong enough signals, $\boldsymbol\theta$ are expected to  have a
Gaussian distribution centered around the true values, with
covariance matrix 
$\tn{Cov}^{ab}=(\Gamma^{-1})^{ab}$, where $\Gamma ^{ab}$ is the Fisher 
information matrix which contains the partial derivatives of the template with 
respect to the binary parameters \cite{PW95}.

To model the waveform we use the TaylorF2 approximant in the frequency 
domain, assuming the stationary phase approximation $h(f)={\cal A}(f)e^{i\psi(f)}$
\cite{DIS00}.
The post-Newtonian expansion of the phase includes spin-orbit and tidal 
corrections. It can be written as  $\psi(f)=\psi_{\tn{PP}}+\psi_{\tn{T}}$, i.e. a 
sum of a point-particle term (see \cite{BFIJ02,BDE04} for the complete expression) 
and a tidal contribution \cite{VFH11,DNV12}. The latter is given by 
\begin{equation}\label{Tphase}
\psi_{\textnormal{T}}=-\frac{117\Lambda}{8}
\frac{(1+q)^2}{qm^5}x^{5/2}\Bigg[1+\frac{3115}{1248}x-\pi x^{3/2}+\left(\frac{23073805}{3302208} +\frac{120}{81}\right)x^{2}
-\frac{4283}{1092}\pi x^{5/2}\Bigg]\ , 
\end{equation} 
where $x=(m \pi f)^{5/3}$, $m=M_{\tn{BH}}+M_{\tn{NS}}$ is the total mass of the system, 
and $\Lambda$ is the averaged tidal deformability, which for BH-NS binaries 
reads \cite{FH08}: $\label{lambda} \Lambda=\lambda_{2}\frac{\left(1+12q\right)}{26}$.

We consider non-rotating NSs, as this is believed to be a reliable approximation 
of real astrophysical systems \cite{BC92,K92}. 
Therefore, our template is  fully specified by 6 parameters\footnote{The 
signal amplitude $\ln {\cal A}$ is uncorrelated with the other variables, so we 
perform derivatives only with respect to the remaining parameters.} 
$\boldsymbol\theta=(t_{c},\phi_{c},\ln{\cal M},q,\Lambda,\beta)$ 
where $t_{c},\phi_{c}$ are the time and phase at the coalescence and 
$\beta$ is the 2 PN spin-orbit contribution in $\psi_{\tn{PP}}$. 
We choose the BH spin  aligned with the orbital angular momentum. 
Moreover, since ${\chi}_{\tn {BH}}\leq \vert1\vert$, $\beta \lesssim 9.4$; 
therefore we consider the following prior probability distribution on $\beta$:
$p^{(0)}(\beta)\propto \tn{exp}\left[-\frac{1}{2}\left(\beta/
9.4\right)^{2}\right]\ .$

In our analysis we consider both second and 
third generation detectors. For AdvLIGO/Virgo we use the \texttt{ZERO\_DET\_high\_P} 
noise spectral density of AdvLIGO \cite{zerodet}, in the frequency
ranges $[20\ \textnormal{Hz},f_{\textnormal{ISCO}}]$;
for the Einstein Telescope we use  the analytic fit of the  sensitivity
curve provided in \cite{ET2}, in the range 
 $[10\ \textnormal{Hz},f_{\textnormal{ISCO}}]$.
$f_{\tn{ISCO}}$ is the frequency of the Kerr ISCO including
corrections due to NS self-force \cite{Fav11}.

We model the NS structure by means of piecewise polytropes, 
 \cite{RMSUCF09}. Indeed we consider four EoS, labeled as \texttt{2H},\texttt{H},
\texttt{HB} and \texttt{B}, which denote very stiff, stiff, moderately stiff and soft 
nuclear matter, respectively.  The  stellar parameters  for 
$M_{\tn{NS}}=(1.2,1.35)M_{\odot}$, are shown in Table~\ref{Tableconf}.

\begin{table}[h!]
\centering
\begin{tabular}{c|ccc||cccc}
\hline
\hline
\texttt{EoS} & $M_{\tn{NS}} (M_{\odot})$ & ${\cal C}$  
& $\lambda_{\tn{2}}$ (km$^5$) & $M_{\tn{NS}} (M_{\odot})$ & ${\cal C}$  
& $\lambda_{\tn{2}}$ (km$^5$)\\
\hline
 \texttt{2H} & 1.2  & 0.117  & 75991 & 1.35  & 0.131  & 72536 \\
 \texttt{H}   & 1.2  & 0.145  & 21232 & 1.35  & 0.163  & 18964\\
 \texttt{HB} & 1.2  & 0.153  & 15090 & 1.35  & 0.172  & 13161\\
 \texttt{B}   & 1.2  & 0.162  & 10627 & 1.35  & 0.182  & 8974\\
 \hline
 \hline
\end{tabular}
\caption{For each EoS we show the NS mass, the compactness 
${\cal C}=M_{\tn{NS}}/R_{\tn{NS}}$, and the tidal 
deformability $\lambda_{2}$.}
\label{Tableconf}
\end{table}


\section{Numerical results}

Following the strategy previously outlined,
we compute the minimum and maximum probabilities (\ref{Pminmax}) 
that the coalescence of a BH-NS system produces a 
remnant disk with mass above a threshold  $\bar{M}_{\tn{rem}}$, 
for the NS models listed in Table~\ref{Tableconf} 
and different values of the mass ratio $q$. The results are given in 
Table~\ref{table2}-\ref{table3}, for $q=3$ and $q=7$, black hole spin 
$\chi_{\tn{BH}}=(0.2,0.5,0.9)$, $M_\tn{NS}=(1.2,1.35)~M_\odot$, and 
disk mass thresholds $\bar{M}_{\tn{rem}}=0.01M_{\tn{NS}}$.

For AdvLIGO/Virgo we put the source at a distance of $100$ Mpc.
For ET the binary is at $1$ Gpc. In this case the signal must be
suitably redshifted \cite{CF94,MGF13}, and  we have assumed that $z$ is known
with a fiducial error of the order of 10\% \cite{MR12}.


\begin{table*}[ht]
\centering
\begin{tabular}{cc|ccc|ccc}
&AdV&& $q=3$ &&& $q=7$ \\
\hline
& $M_{\tn{NS}}=1.2M_{\odot}$&& $\chi_{\tn{BH}}$ &   & & $\chi_{\tn{BH}}$  \\
\hline
\texttt{EOS} & ${\cal C}_{\lambda}$ & $0.2$ 
& $0.5$ & $0.9$ & $0.2 $ & $0.5$ & $0.9$\\
\hline
\texttt{2H} & 0.118 & 1 & 1 & 1  & 0.4 & [0.8-0.9]  &  1  \\
\texttt{H}  &  0.147 & [0.6-0.9] & 1 & 1 & 0.4 & 0.4  & [0.8-0.9] \\
\texttt{HB} & 0.155 & [0.5-0.7] & [0.9-1] & 1 & 0.4 & 0.4 & [0.7-0.8]  \\
\texttt{B}  &  0.164 & [0.4-0.6] & [0.7-0.8] & 1 & 0.4 & 0.4 & [0.6-0.7]  \\
\hline
& $M_{\tn{NS}}=1.35M_{\odot}$ &\\
\hline
\texttt{2H} & 0.132 & 1 & 1 & 1  & 0.3 & [0.4-0.5] & 1  \\
\texttt{H}  & 0.164 & [0.4-0.6] & [0.8-0.9] & 1  & 0.4 & 0.4  & 0.7  \\
\texttt{HB} & 0.173 & [0.4-0.5] & [0.6-0.8] & 1  & 0.4 & 0.4 & 0.6   \\
\texttt{B}  & 0.184   & [0.4-0.5] & [0.5-0.6] & 1  &  0.4  & 0.4 & 0.5  \\
\end{tabular}
\caption{We show the probability range [P$_{\tn{MIN}}$,P$_{\tn{MAX}}$]
that the coalescence of a BH-NS binary produces a disk mass larger than 
$\bar{M}_{\tn{rem}}=0.01M_{\tn{NS}}$ for AdLIGO/Virgo (AdV), for binaries 
with $q=3$ and $q=7$, NS masses (1.2,1.35)$M_{\odot}$, and BH spin 
$\chi_{\tn{BH}}=(0.2,0.5,0.9)$. Sources are assumed to be at $d=100$ Mpc. 
The star compactness ${\cal C}_{\lambda}$ is estimated throughout the 
universal relation (\ref{Clambda}).}\label{table2}
\vspace{0.1cm}
\begin{tabular}{cc|ccc|ccc}
&ET&& $q=3$ &&& $q=7$ \\
\hline
& $M_{\tn{NS}}=1.2M_{\odot}$&& $\chi_{\tn{BH}}$ &   & & $\chi_{\tn{BH}}$  \\
\hline
\texttt{EOS} & ${\cal C}_{\lambda}$ & $0.2$ 
& $0.5$ & $0.9$ & $0.2 $ & $0.5$ & $0.9$\\
\hline
\texttt{2H} & 0.118  & 1 & 1 & 1  & [0.3-0.4] & [0.7-0.8] & 1 \\
\texttt{H}  &  0.147  & [0.9-1] & 1 & 1  & 0.3 & [0.3-0.4] & 1\\
\texttt{HB} & 0.155 &  [0.7-0.8] & 1 & 1  & 0.3 & 0.3 & 0.9 \\
\texttt{B}  &  0.164  & [0.5-0.6] & 1 & 1  & 0.3 & 0.3 & [0.7,0.8] \\
\hline
& $M_{\tn{NS}}=1.35M_{\odot}$ &\\
\hline
\texttt{2H} & 0.132  &  1 &1 &1  & 0.2 & [0.4,0.5] &1 \\
\texttt{H}  & 0.164   &  [0.4-0.6] & 1 &1 & 0.3 & 0.3 & 0.8 \\
\texttt{HB} & 0.173  &  [0.3-0.4] & 0.8 &1 & [0.3-0.4] & [0.3-0.4] & 0.6 \\
\texttt{B}  & 0.184   &  [0.3-0.4] & 0.6 &1 & 0.4 & 0.4 & [0.5-0.6] \\
\end{tabular}
\caption{Same of Table~\ref{table2}, but for the Einstein Telescope (ET). 
In this case we assume prototype BH-NS binaries at $d=1$ Gpc.}
\label{table3}
\end{table*}

The first clear result is that as the BH spin approaches the highest value 
we consider, $\chi_{\tn{BH}}=0.9$,  and for low mass ratio $q=3$,
the probability that a BH-NS coalescence produces a disk with
mass above the threshold  is insensitive to the NS
internal composition, and it approaches  unity for all 
considered configurations. These would be good candidates for
GRB production.
For the highest mass ratio we consider,  $q=7$, the probability to form a sufficiently 
massive disk depends on the NS mass and EoS, and on the detector.
In particular, it decreases as the EoS softens, and as the NS mass
increases. This is a general trend, observed also for smaller values of
$\chi_{\tn{BH}}$.   
However, when $\chi_{\tn{BH}}=0.9$ the probability that the coalescence
is associated to a SGRB is always $\gtrsim 50\%$ .

Let us now consider  the results for $\chi_{\tn{BH}}=0.2$.  
If the NS mass is $1.2~M_\odot$ the probability that a detected GW
signal from a BH-NS coalescence is associated to the formation of a
black hole with a disk of mass above threshold is $\gtrsim 50\%$ for
both AdvLIGO/Virgo and ET, provided $q=3$. For larger  NS mass,
this remains true only if the NS equation of state is stiff (\texttt{2H}
or \texttt{H}).
High values of $q$ are disfavored.

When the  black hole spin has an intermediate value, say
$\chi_{\tn{BH}}=0.5$,  Table~\ref{table2} shows that, 
the NS compactness plays 
a key role in the identification of good candidates for GRB production,
for both detectors. Again large values of the mass ratio yield small
probabilities.

The range of compactness shown in  Table~\ref{table2}-\ref{table3}
includes neutron stars with radius ranging within $\sim [10-15]$ km.
From the table it is also clear that if we choose a
compactness smaller than the minimum value,
the probability of generating a SGBR increases, and
the inverse is true if we consider compactness larger than our maximum.


\section{Conclusions}

The method developed in this paper can be used in several different ways.
In the future, gravitational wave detectors are expected to reach a
sensitivity sufficient to extract the parameters on which our analysis
is based, i.e., chirp mass, mass ratio, source distance, spin and tidal
deformability. We can also expect  that the steady improvement of the 
efficiency of computational facilities experienced in recent years will
continue, reducing the time needed to obtain these parameters from a
detected signal. Moreover, the higher sensitivity  
will allow us to detect sources in
a much larger volume space, thus increasing the detection rates.
In this perspective, the method we envisage in this paper will be useful
to trigger the electromagnetic follow-up of a GW detection, 
searching for the afterglow  emission of a SGRBs.

Until then, the method we propose can be used 
in the data analysis of advanced detectors as follows:
\begin{itemize}
\item
Table \ref{table2}-\ref{table3} indicate the systems that are more likely to
produce accretion disks sufficiently massive to generate a SGRB.
The table can be enriched including more 
NS equations of state or more binary parameters;
however, it already contains a clear information on which is
the range of  parameters to be used in the GW data analysis,
if the goal is  to  search for  BH-NS signals which may be associated to a
GRB. For instance, 
Table \ref{table2}-\ref{table3} suggests that searching for mass ratio smaller 
than, or equal to, $3-4$, and values of the black-hole angular momentum 
larger than $0.5-0.6$ would allow us to save time and computational 
resources in low-latency search. In addition, it would allow us to gain 
sensitivity in externally triggered searches performed in time coincidence 
with short GRBs observed by gamma-ray satellites.
\item
If a SGRB is observed sufficiently close to us in the electromagnetic
waveband, the parameters of the GW signal detected in coincidence would
allow us to set a threshold on the mass of the accretion disk.  If the GW
signal comes, say, from a system with a BH with spin
$\chi_{\tn{BH}}=0.5$, mass ratio $q=7$, and neutron star mass
$M_{\tn{NS}}=1.2 M_\odot$, from Table~\ref{table2},  equations of state
softer than the EoS \texttt{2H} would be disfavored.  Thus, we would be
able to shed light on the dynamics of the binary system, on its
parameters and on the internal structure of its components. We would
enter into the realm of gravitational wave astronomy.  
\end{itemize}
Finally,  it is worth stressing that as soon as the fit 
(\ref{torusFit}) is extended to NS-NS coalescing
binaries, this information will be easily implemented in
our  approach.
With the rate of NS-NS coalescence higher than that of BH-NS, our
approach will acquire more significance, and will
be a very useful tool to study these systems.



\begin{thebibliography}{99.}
\bibitem{LIGOVirgo} \url{http://www.ligo.caltech.edu}, \url{http://www.ego-gw.it}.
\bibitem{ET1} \url{http://www.et-gw.eu}.
\bibitem{LR07} William~H.~Lee and Enrico~Ramirez-Ruiz, New J. Phys. {\bf 9}, 17 (2007).
\bibitem{Abadieb} J.~Abadie {\it et al.}  (LIGO Scientific Collaboration), 
Astrophys.\ J.\  {\bf 755}, 2 (2012).
\bibitem{Abadiec} J.~Abadie {\it et al.}  (LIGO Scientific Collaboration), 
Astrophys.\ J.\  {\bf 760}, 12 (2012).
\bibitem{LIGOVirgoplan} J.~Aasi et al., arXiv:1304.0670 (2013).
\bibitem{Abadiea} J.~Abadie {\it et al.}  (LIGO Scientific Collaboration), 
Astron. Astrophys.  {\bf 541} A155 (2012).
\bibitem{F12} Francois Foucart, Phys. Rev. {\bf D 86}, 124007 (2012).
\bibitem{GPRTL13} B.~Giacomazzo, R.~Perna, L.~Rezzolla, E.~Troja and 
D.~Lazzati, Astrophys. J. Lett., {\bf 762}, L18 (2013).
\bibitem{BPT72} J.~M.~Bardeem, W.~H.~Press, and S.~A.~Teukolsky, Astrophys. 
J. {\bf 178}, 347 (1972).
\bibitem{KOST11} K.~Kyutoku, H.~Okawa, M.~Shibata, and K.~Taniguchi, Phys. Rev. D {\bf 84}, 
064018 (2011).
\bibitem{ELSB09} Z.B.~Etienne, Y.T.~Liu, S.L.~Shapiro and T.W.~Baumgarte, Phys. Rev. D {\bf 79}, 044024 (2009).
\bibitem{FDKSST12} F.~Foucart, {\it et al.}, Phys. Rev. D {\bf 85}, 044015 (2012).
\bibitem{FDKT11} F.~Foucart, M.D.~Duez, L.E.~Kidder, and S.A.~Teukolsky, Phys. Rev. D {\bf 83}, 024005 (2011).  
\bibitem{Piran04}  T.~Piran,  Rev.\ Mod.\ Phys.\  {\bf 76}, 1143 (2005).
\bibitem{SLB13} N.~Stone, A.~Loeb, and E.~Berger, Phys. Rev. {\bf D}, 084053 (2013).
\bibitem{MCFGP13}  A.~Maselli, V.~Cardoso, V.~Ferrari, L.~Gualtieri and P.~Pani, Phys.\ Rev.\ D {\bf 88}, 023007 (2013).
\bibitem{DABV13} W.~Del Pozzo, {\it et al.}, Phys.\ Rev.\ Lett. {\bf 111}, 071101 (2013).
\bibitem{DNV12} T.~Damour, A.~Nagar, and L.~Villain, Phys. Rev. {\bf D} 85, 123007 (2012).
\bibitem{Retal13} J.~Read, {\it et al.}, Phys. Rev. {\bf D} 88, 044042 (2013).
\bibitem{PROR11} F.~Pannarale, L.~Rezzolla, F.~Ohme, and J.~S.~Read, Phys. Rev. {\bf D} 84, 104017 (2011).
\bibitem{MGF13} A.~Maselli, L.~Gualtieri, and V.Ferrari, Phys. Rev. {\bf D} 88, 104040 (2013).
\bibitem{RMSUCF09} J.~Read, {\it et al.}, Phys. Rev. {\bf D} 79, 124033 (2009).
\bibitem{PW95} E.~Poisson and C.~M.~Will, Phys. Rev. D {\bf 52}, 2 (1995).
\bibitem{DIS00} T.~Damour, B.~Iyer, and B.~Sathyaprakash, Phys. Rev. D {\bf 62} 084036 (2000).
\bibitem{BFIJ02} L.~Blanchet, G.~Faye, B.~R.~Iyer, and B.~Jouget, Phys. Rev. D {\bf 65}, 061501 (2004).
\bibitem{BDE04} L.~Blanchet, T.~Damour, G.~E.~Farese, and B.~R.~Iyer, Phys. Rev. Lett. 
{\bf 93}, 091011 (2004).
\bibitem{VFH11} J.~Vines, E.E.~Flanagan, and T.~Hinderer, Phys. Rev. {\bf D} 83, 084051 (2011).
\bibitem{FH08} E.E.~Flanagan and T.~Hinderer, Phys. Rev. {\bf D} 77, 021502 (2008).
\bibitem{BC92} L.~Bildsten and C.~Cutler, Astrophys. J {\bf 400}, 175 (1992). 
\bibitem{K92} C.~S.~Kochanek, Astrophys. J {\bf 398}, 234 (1992). 
\bibitem{zerodet} D.~Shoemaker, \texttt{https://dcc.ligo.org/cgi-bin/DocDB/\\ShowDocument?docid=2974}.	
\bibitem{ET2} B.S. Sathyaprakash  and B.F.~Schultz, Living Rev. Relativity {\bf 12}, 2 (2009).
\bibitem{Fav11} Marc Favata, Phys. Rev. D {\bf 83}, 024028 (2011).
\bibitem{CF94} C.~Cutler and E.~E.~Flanagan, Phys.\ Rev.\ D {\bf 49} (1994) 2658.
\bibitem{MR12} C.~Messenger and J.~Read, Phys.\ Rev.\ Lett.\  {\bf 108} (2012) 09110.
\end{thebibliography}
\end{document}